\documentclass{llncs}

\usepackage{amssymb}
\usepackage{graphicx}
\usepackage{xcolor}

\usepackage{dsfont}
\usepackage{tabularx}
\usepackage{tabulary}

\usepackage{url}
\urldef{\mailsa}\path|{mchgoya, pcaballe}@ull.es|    

\usepackage{tabularx}
\usepackage{subfigure}
\usepackage{dsfont}

\usepackage{multirow}

\begin{document}

\pagestyle{headings}  

\title{Analysis of Lightweight Cryptographic Solutions for Authentication in 
IoT
\thanks{Research supported by  MINECO  and  FEDER  under Project TIN2011- 25452.}}

\titlerunning{Analysis of Lightweight Cryptographic Solutions for Authentication in the
IoT}

\author{C. Hern\'andez-Goya, P. Caballero-Gil%
}

\authorrunning{C. Hern\'andez Goya et al.}
\institute{Department of Statistics, O.R. and
Computing. University of La Laguna. 
Spain. 
\mailsa\\}

\maketitle

\begin{abstract}
Currently, special attention is being paid to scenarios where the
interconnection of devices with heterogeneous computational and communication
capabilities it is required. It is essential to integrate security services during the stages of
design and deployment of these networks since many of these scenarios provide critical services
such as medical health, payment systems, military affairs, access control, e-banking, etc.  This
work analyses several cryptographic primitives related to entity authentication providing robust
solutions according to device capabilities.
\end{abstract}

\section{The Internet of Things (IoT) setting}

The Internet of Things (IoT) \cite{vermesan2009internet} may be defined as a
   ``... \textbf{dynamic} global network infrastructure with \textbf{self configuring capabilities} based on standard
and interoperable communication protocols where physical and virtual ''things`` have \textbf{identities},
physical attributes, and virtual personalities, use intelligent interfaces, and are seamlessly
integrated into the information network.'' 

This paradigm has become a reality thanks to the integration of devices such as smartphones, PDAs,
sensor devices and RFID tags in the daily routine of users. Ideally, IoT allows people and
things to be connected Anytime, Anyplace, with Anything and Anyone, using Any path/network and Any
service. This philosophy poses many challenges not only into the integration of different
technologies, but also when trying to provide IoT with security services. For example, a highly
dynamic nature forces to consider proactive protocols, wireless communication may ease
eavesdropping, dedicated devices are integrated increasing implementation difficulties and 
scalability problems arise due to the large number of devices that may be interconnected. Besides
all these potential problems, it is not possible to rely on a fixed and centralized
infrastructure to alleviate management security tasks.    

Although there are many security services to integrate in this environment this paper
focuses on entity authentication. Ensuring that the information flow is established
between legitimate devices is a major concern in critical scenarios. 

Section \ref{Sec:InsOutsAuthentication} is devoted to the description of different solutions to the
authentication problem remarking their limitations and weaknesses for the IoT setting. The
definition and benefits that Zero-Knowledge Proofs (ZKPs) provide are described there as well. Some
basic concepts regarding Elliptic Curve Cryptography, needed to understand the protocols proposed,
are included in section \ref{Sec:LightweightCryptography}. Section \ref{Sec:ZKPIOT} gathers the
proposal of three authentication algorithms and their comparison. Last section is devoted to the
conclusions drawn from this work.

\section{Ins and outs of authentication}
\label{Sec:InsOutsAuthentication}
The goal of entity authentication protocols is to allow a verifier (B) to gain assurance that the
identity of a prover (A) is as declared. Conventional cryptographic solutions generally require
specific infrastructure and messages exchange among nodes. Just transferring these solutions to
the IoT setting is not a correct approach \cite{Bab:Mah:Sta10}. Solutions devised to sensor
networks \cite{Eschenauer:2002:KSD:586110.586117} rely on the presence of particular nodes acting as
gateways to connect with Internet. Communication technologies behind IoT (e.g. RFID, NFC,
WiFi-Direct, etc.) do not include either robust solutions. For example, in some cases the length of
the keys is inappropriate \cite{Jue:06}.

Albeit there are many restrictions to consider there are some general issues that may be
established as rule of thumb when designing authentication protocols for this
setting. First, the computational steps and their cost must be kept to a minimum in order to save
battery power. Besides, the characteristics of the devices should be considered in order to 
distribute the computational burden among nodes according to their capabilities. Finally, the number
of exchanged messages and their length are parameters that should be chosen with proper
care. Below we have classified and analyzed general authentication tools grouping them according to
the cryptographic paradigm they use.

\subsection{Password-based methods}
\label{Sec:Password-basedMethods}
There are many impairments that discourage the implementation of password-based methods in the IoT:
secure storage requires integrating key management services and passwords length may hinder
usability. However, scalability problem is the biggest deterrent against using this
approach since the number of devices that a user is responsible for can convert
password management into a nightmare.

One-time passwords improve protection against eavesdropping and replay attacks since non-stationary
information is used. However, they do not represent the ideal solution since password redefinition
requires either shared lists or sequential updating through one way functions. The first method is
difficult to adapt to dynamic and decentralized networks (one-to-one shared password lists are
necessary). Additionally, when synchronization fails replay attacks became an actual possibility.

\subsection{Challenge-response alternatives}
In general, the challenge-response approach makes possible to demonstrate prover's knowledge about a
secret associated with her, in order to be authenticated, without revealing it. It is assumed
certain level of trust among parties. 

These methods may be implemented based on secret key demanding then additional tools  
to define the shared key. In this case, procedures for securely updating and storing
keys may affect scalability.

\subsection{Zero Knowledge Proofs: the selected alternative}
Zero Knowledge Proofs (ZKPs) contains challenge-response technique as building block. The main
difference with the conventional challenge-response approach is that the objective there is to prove
knowledge of a shared secret while in ZKPs sharing secrets among participants is completely avoided.
ZKPs allow an entity A (prover) to prove the possession of certain secret without
revealing anything about it to another entity B (verifier). The basic outline for one iteration of
these protocols is composed by three steps described below.

\begin{enumerate}

   \item \textbf{Witness}: $A$ chooses a random element from a predefined set, keeping it secret
(the compromise). From this element, $A$ generates a piece of information
(the witness) and transfers it to B. Normally, the witness corresponds to a random instance of the
base problem. 
   \item \textbf{Challenge}: $B$ randomly chooses  one question that should be answered by $A$. The
question is related to the compromise and also to the secret associated to $A$'s credentials.
   \item \textbf{Response}: $A$ sends the answer corresponding to the challenge chosen by $B$. 
 \end{enumerate}

The iteration concludes when $B$ has checked the answer given by $A$.

 From the previous description it may be appreciated its interactive behaviour, the
characteristics of the three main messages exchanged between participants and its probabilistic
nature. \footnote{There is certain negligible probability for $A$ to answer correctly the questions
posed by $B$ without knowing the secret associated to her identity. It may be reduced by increasing
the number of iterations or defining the challenges on a set  with a suitable
cardinality.} These protocols are part of the standard ISO 9798-5 devoted to entity authentication. 

%

Some of the main advantages of using ZKPs protocols in authentication are:
 \begin{itemize}
    \item No secret information is degraded with usage, facilitating identity management
and helping to avoid scalability problems.
    \item The encryption of the exchanged messages is not necessary. It leads to savings in
demanding operations.
    \item It is possible to improve efficiency in several ways. Two of them are included in the
proposals: 
      \begin{itemize}
       \item Optimizing the consumption of bandwidth and computational resources by using
Elliptic Curve Cryptography (ECC). 
       \item Reducing the iterations defining the challenges from a set with enough
cardinality.  
      \end{itemize}

      \item It is possible to reach mutual authentication by modifying the basic protocol. This property is interesting because the communication architecture in many scenarios of the IoT,  communication between devices emulates the peer-to-peer paradigm. 
    \item Probably one of the best characteristic in ZKPs is their application when there is a complete lack of trust between participants.
    \item The participation of third parties is not required. Although devices may have direct access to Internet, assuming the availability of a centralized service is too optimistic, specially when the scenario involves emergency situations. 
  \end{itemize}

In the RFID environment ZKPs has been presented as a promising alternative when dealing with  privacy issues. \cite{engberg2004zero} proposes an identification scheme for RFID tags specifically based on ZKPs that was implemented as part of the project Zeroleak. However, this approach is not based on the original zero knowledge concept since the objective there is proving the knowledge of a secret shared key. There are also several works where
conventional ZKPs are proposed for RFID \cite{Liu:Nin:11},
\cite{Chatzigiannakis:11}, \cite{DBLP:conf/secrypt/Hutter09} and NFC \cite{Alpar:12} technologies. Generally, the ZKPs proposed for the IoT, including the protocols described in this paper,  use  ECC to optimize their deployment \cite{Ram:Aro:11}. Next section deals with the basic principles of ECC and, more precisely, with the Elliptic Curve Discrete Logarithm Problem (ECDLP) since it is the base problem in the protocols presented here.

\section{Lightweight cryptography basics}
\label{Sec:LightweightCryptography}

Before detailing the authentication protocols, this section introduce just the fundamental theory related to ECC, indispensable to describe them. The main advantages of using ECC are the greater computational complexity of problems and the smaller key length required for a particular security level. A complete guide on the subject may be found in \cite{Hankerson:2003}.

The Weierstrass equation defines an elliptic curve over a finite field $F_{p}$ (p a prime number) as:  $E(\mathds{F}_p) =  \{(x, y) \in \mathds{F}_p \times \mathds{F}_p: y^2 = x^3 + ax + b\} \cup \{O\}$ where  $a, b \in \mathds{F}_p, 4a^3 + 27 b^2 \neq 0$ and $O$ represents the point at the infinity. When using $p =2$ the implementation in the devices will become easier.  Considering the finite set of points on the curve $E(\mathds{F}_p)$, the addition of two points and the multiplication of a point by an integer, the group   $<E(\mathds{F}_p),+>$ is defined. 
%
%
%

Now it is possible to introduce the Elliptic Curve Discrete Logarithm Problem (ECDLP). Given this group, a base point P (a generator of a cyclic subgroup G of $E(\mathds{F}_p)$)  and a second point $Q$, solving the ECDLP consist of finding the unique integer $k$, $0 \leq k \leq m - 1$ such that $Q =  k \ast P$. In the ZKPs included in next section $(Q,P)$ corresponds to the prover's public identification while $k$ is the secret identification.

\section{ZKP for the IoT}
\label{Sec:ZKPIOT}
Three different non-iterative ZKPs  built on the ECDLP are included in this section. In order to achieve bandwidth optimization, the number of iterations  has been reduced by using  a random integer  or a hash function as challenge. In that way, challenges are defined in a set with enough cardinality to assure that the cheating probability in one iteration is negligible.

\subsection{One-way authentication}

The first ZKP (ZKP1) is described in table 1. The number of iterations to carry out is  only one because challenges are defined through a private and robust cryptographic hash function selected by the verifier. Since the computational and communication requirements are not excessive, this protocol is suitable to authenticate RFID devices and also devices belonging to low-rate wireless personal area network, LR-WPAN.

  \begin{table}[!ht]
 \begin{scriptsize}
  \begin{center}
  \begin{tabularx}{\textwidth}{|X|X|}
   \hline
   \centering{\textbf{Stages} }& \textbf{ZKP1} \\ \hline
  \centering{Bootstrapping} &$p$ prime number, $E$ elliptic curve in ${Z}_p$, $P \in E$\\\hline
  \centering{$A's$ secret identification} & $a \in {Z}_p$ \\\hline
  \centering{$A's$ public identification: \centering $PuId_A$ } & $a\ast P \in E$ \\\hline
  \centering Compromise: &   \\ 
  \centering $A's$ secret & $x\in_r {Z}_p $ \\\hline
  \centering Witness:  &   \\ 
  \centering $A \rightarrow B$ & $w = x \ast P \in E$ \\\hline
  \centering Challenge: &    \\ 
   \centering $A \leftarrow B$  & $e = hash(P,a\ast P, x\ast P)$  \\\hline 
 \centering Answer: &    \\ 
 \centering $A \rightarrow B$  & $y = x + a \ast e \in {Z}_p $  \\\hline 
  \centering Verification: &    \\ 
  \centering {B checks}& $y\ast P - e\ast PuId_A = w$ \\\hline
 \end{tabularx}
 \end{center}
 \end{scriptsize}
  \label{tab2:ZKP1} \caption{One-way authentication: ZKP1}
  \end{table}
Table 2 contains another proposal for one-way authentication. Here, on-line computational requirements are reduced since the compromises are calculated and loaded into the entity previously to the protocol execution. Hence, this version is advisable for environments where devices have asymmetric computational capabilities. An example is RFID technologies where tags are mostly passive. Again, just one iteration is required. In this case, challenges are defined from a set with enough cardinality.

\begin{table}[!ht]
 \begin{scriptsize}
 \begin{center}
 \begin{tabularx}{\textwidth}{|X|X|}
  \hline
   \centering{\textbf{Stages} }& \textbf{ZKP2}\\ \hline
 	\centering{Bootstrapping} &$p$ prime number, $E$ elliptic curve in ${Z}_p$, $P \in E$ \\\hline
 	
 	\centering{$A's$ secret identification} &  $a \in {Z}_p$ \\\hline
 	\centering{$A's$ public identification: \centering $PuId_A$ } & $a\ast P \in E$ \\\hline
 	
 	\centering Compromise: &   \\ 
 	\centering $A's$ secret & $\{x_1\ast P, x_2\ast P,...,x_n \ast P\} \in E$, with \ $x_i\in_r Z_p$ \\\hline
 	
 	\centering Witness:  & \\ 
 	\centering $A \rightarrow B$ & $ w = hash(x_j \ast P + x_k \ast P)$, with $ j,k \in_r \{1,
 2,...,n\} $ \\\hline
 	
 	\centering Challenge: & \\ 
 	\centering$A \leftarrow B$  & $e \in_r {Z}_p $ \\ \hline
 
 	\centering Answer: & \\ 
 	\centering $A \rightarrow B$  & $y = x_j + x_k -a \ast e \in{Z}_p$ \\\hline 
 
 	\centering Verification: & \\ 
 	\centering {B checks}&  $hash(y\ast P + e\ast PuId_A) = w$ \\\hline
 
 \end{tabularx}
 \end{center}
 \end{scriptsize}
 \label{tab3:ZKP2} \caption{One way authentication: ZKP2}
 \end{table}
\subsection{Mutual authentication and key agreement at once}
The last ZKP (table 3) provides mutual authentication and makes possible establishing a shared session key simultaneously. The protocol becomes an Authentication and Key Agreement (AKA) protocol completely symmetric for both participants. Hence, it is specially suitable for NFC technology. Here challenges are robustly defined by applying a hash function on information provided by both users. The result of this operation may be used as shared key in order to provide confidentiality users communication. 

It should be pointed out that the challenge is not transferred, both devices calculate it independently by using its secret identification, the public identification of the other party, the witness and its own compromise. All this information is the input of a hash function and the result obtained, which is exactly the same at both ends of the communication, is used both as challenge and as  shared secret key.


Considering how the shared key is defined, it makes sense to find out if the protocol holds forward security. This property is widely sought in AKA protocols. It consists of guaranteeing that even when the private key of one or more parties is compromised, previous session keys are not affected. The key is established not only depending on the secret keys of users but also on the compromises that are randomly and independently defined in each execution what makes it more robust. In order to obtain the shared key, the eavesdropper should gather too much information: one of the compromises and the corresponding entity secret key. The other alternative is trying to obtain it from the equation associated to the verification. In this case, the opponent will have to solve an ECDLP instance, what is computational infeasible.

 \begin{table}[!ht]
 \begin{scriptsize}
 \begin{center}
 \begin{tabularx}{\textwidth}{|X|X|}
  \hline
   \centering{\textbf{Stages} } & \textbf{ZKP3}\\ \hline
 	\centering{Bootstrapping} & $p$ prime number, $E$ elliptic curve in ${Z}_p$, $P \in E$\\\hline
 	\centering{$A's$ secret identification} & $a \in {Z}_p$ \\
 	\centering{$B's$ secret identification} & $b \in {Z}_p$ \\\hline
 
 	\centering{$A's$ public identification: \centering $PuId_A$ }  &$a\ast P \in E$  \\
 	\centering{$B's$ public identification: \centering $PuId_B$}   &$b\ast P \in E$ \\\hline
 
 	\centering Compromise: & \\ 
 	\centering $A's$ secret & $x_A\in_r {Z}_p $ \\
  	\centering $B's$ secret & $x_B\in_r {Z}_p $\\\hline
 
 	\centering Witness:  & \\ 
 	\centering $A \rightarrow B$ & $w_A = x_A \ast P \in E$ \\
 	\centering $A \leftarrow B$ & $w_B = x_B \ast P \in E$  \\\hline
 
 	\centering Challenge: & \\ 
 	\centering $A$ and $B$ compute  & $e = hash(P,a\ast b \ast P, x_A \ast x_B \ast P) $\\ \hline
 
 	\centering Answer: & \\ 
 	\centering $A \rightarrow B$ & $y_A = x_A + a \ast e \in {Z}_p $ \\ 
 	\centering $A \leftarrow B$  & $y_B = x_B + b \ast e \in {Z}_p $  \\\hline
 	
 	\centering Verification:  &   \\ 
 	\centering {B checks} & $y_A\ast P - e\ast PuId_A = w_A$\\
 	\centering A checks &  $y_B\ast P - e\ast PuId_B = w_B$ \\\hline
 
 \end{tabularx}
 \end{center}
 \end{scriptsize}
 \label{tab4:MutualZKPAKA}\caption{Mutual authentication and key agreement}
 \end{table}

\subsection{Comparing the proposals}
The parameters used for the comparison of the ZKPs  are: message length (communication), number of products involving a scalar and a point (computation) and bits required to store the secrets (memory) (table \ref{tab5:ZKPsComparison}). The comparison referred to bandwidth consumption has been made assuming that the hash function used is the  new standard SHA-3 \cite {Kavun:2010} and the elliptic curve used is P-192 \cite {Ros:Zav:11}. These choices have been made in order to obtain conclusions about the performance of the proposal independently of the platform where they are used. In order to estimate the computing requirements and to avoid dependency on the used technology, $ \lambda $ stands for the cost of a product operation and $ \mu $ for the hash computation. The needs of both participants are distinguished since their computational capabilities may be significantly different.

\begin{table}[!ht]
\begin{scriptsize}
\begin{center}
 \begin{tabularx}{\textwidth}{|X|X|X|X|X|X|}
\hline
 \multicolumn{2}{|c|}{\textbf{ZKP1}} & \multicolumn{2}{c|}{\textbf{ZKP2} }& \multicolumn{2}{c|}{\textbf{ZKP3}}  \\ \hline

\multicolumn{6}{|c|}{Bandwidth (transferred bits)} \\ \hline
A & B & A & B &  A & B \\ \hline
 1728 & 128 & 320  & 192 & 1728 & 1728 \\ \hline

\multicolumn{6}{|c|}{Memory (stored secret bits )} \\ \hline 
\multicolumn{2}{|c|}{A} & \multicolumn{2}{c|}{B} &  A & B \\ \hline
\multicolumn{2}{|c|}{384} & \multicolumn{2}{c|}{$n*1536 + 192$} & 384 & 384\\ \hline
  
\multicolumn{6}{|c|}{Computation (significant operations)} \\ \hline
  A & B & A & B & A & B\\ \hline
  $\lambda$ & $2*\lambda + \mu$ & $n*\lambda$ & $2*\lambda +\mu$ & $3*\lambda + \mu$  &
$3*\lambda + \mu$\\\hline
\end{tabularx}

\end{center}
\end{scriptsize}
\label{tab5:ZKPsComparison}\caption{ZKPs Comparison}
\end{table}

From the previous table we may conclude that the second proposal (ZKP2) is the one where less bits are transferred, this is because the witness is defined through a hash function instead of points of the curves. However, the second proposals is the one where more memory is required since the compromises are off-line precomputed  and stored into the tag before any authentication request starts.

\section{Conclusions}
\label{sec:Conclusions}

From the analysis of the authentication tools carried out while developing this work we may state that using ECC is the most efficient way of implementing advanced protocols for resource constrained devices: it outperforms conventional cryptosystems in terms of computation time, memory requirements and energy consumption.

Regarding authentication tools, ZKPs based on ECDLP represent robust and efficient primitives to solve the authentication problem in mobile computing. An important issue is that additional optimization may be included in these protocols to reduce their iterative nature by generalizing the set where challenges are defined. Finally, from the comparison among the methods it may be appreciated that extending ZKPs to get mutual authentication does not convey significant overconsumption of resources.

\bibliographystyle{splncs03}
\bibliography{BiblioIoT}
%
%
%
%
%
\end{document}